\documentclass[aps,prl,reprint,floatfix,superscriptaddress]{revtex4-1}

\usepackage{etoolbox}

\usepackage{amsmath}
\usepackage{amssymb}
\usepackage[utf8]{inputenc}
\usepackage{graphicx}
\usepackage{xcolor}
\usepackage[caption=false]{subfig}
\usepackage{hyperref}
\hypersetup{colorlinks=true,citecolor={blue},linkcolor={blue},urlcolor={blue}}

\newcommand{\mi}{\mathrm{i}}
\begin{document}

\title{Reconstruction of Exciton-Polariton Condensates in 1D Periodic Structures}

\author{S. Yoon}
\affiliation{Center for Theoretical Physics of Complex Systems, Institute for Basic Science (IBS), Daejeon 34126, Korea}

\author{M. Sun}
\affiliation{Center for Theoretical Physics of Complex Systems, Institute for Basic Science (IBS), Daejeon 34126, Korea}
\affiliation{Basic Science Program, Korea University of Science and Technology (UST), Daejeon 34113, Korea}

\author{Y. G. Rubo}
\affiliation{Center for Theoretical Physics of Complex Systems, Institute for Basic Science (IBS), Daejeon 34126, Korea}
\affiliation{Instituto de Energ\'{\i}as Renovables, Universidad Nacional Aut\'onoma de M\'exico, Temixco, Morelos, 62580, Mexico}

\author{I. G. Savenko}
\affiliation{Center for Theoretical Physics of Complex Systems, Institute for Basic Science (IBS), Daejeon 34126, Korea}
\affiliation{Basic Science Program, Korea University of Science and Technology (UST), Daejeon 34113, Korea}

\begin{abstract}
We demonstrate the reconstruction of the exciton-polariton condensate loaded in a single active miniband in one-dimensional microcavity wires with a complex-valued periodic potentials. 
The effect appears due to strong polariton-polariton repulsion and it depends on the type of the single-particle dispersion of the miniband, which can be fine tuned by the real and imaginary components of the potential. 
As a result, the condensate can be formed in a $0$-state, $\pi$-state, or mixed state of spatiotemporal intermittency, depending on the shape of the miniband, strength of interparticle interaction, and distribution of losses in the system.
The reconstruction of the condensate wave function takes place by proliferation of nuclei of the new condensate phase in the form of dark solitons.
We show that, in general, the interacting polaritons are not condensed in the state with minimal losses, neither they accumulate in the state with a well-defined wave vector. 
\end{abstract}

\date{\today}
\maketitle

\emph{Introduction.---} Since the discovery of superfluid--Mott insulator transition with cold atoms in optical lattices~\cite{Jaksch:1998aa,Greiner:2002aa}, the system of bosons in periodic potentials has attracted much attention for both fundamental and applied reasons. While the cold atoms in periodic optical lattices are probably the cleanest system, 
they have extremely low critical temperature due to heavy atomic masses. 
Exciton polaritons (polaritons) in semiconductor microcavities~\cite{Weisbuch:1992aa,Deng:2002aa,Kavokin:2007aa} possess substantially smaller effective masses and can condense not only at liquid Helium~\cite{Kasprzak:2006aa,Balili:2007aa,Lai:2007aa} but also up to the room temperature~\cite{Baumberg:2008aa,Lerario:2017aa}. This makes a system of polaritons in artificial periodic potentials an excellent alternative platform for studying many-body physics, gap solitons~\cite{Tanese:2013aa,Buller:2016aa}, topological polariton states~\cite{Karzig:2015aa,Culevich2016cv}, as well as classical~\cite{Ohadi:2017aa} and quantum~\cite{Liew:2018aa} simulators. 

The physics of polariton condensation is quite different from traditional cold atoms. An external pumping (coherent or incoherent) is required to create and maintain polaritons due to their finite lifetime in the microcavity, which usually prevents the particles to reach thermal equilibrium, so that the steady-state condensate can be formed in an excited state with many-body correlations.
In particular, for polariton condensates in periodic potentials, the condensation at the edges of the Brillouin zone, namely $\pi$-condensation in one-dimensional (1D) lattices~\cite{Lai:2007aa} and $p$- and $d$-condensation in two-dimensional (2D) lattices~\cite{Kim:2011aa}, as well as the mixed condensates~\cite{ZhangE1516} have been observed. 

Loading cavity photons and quantum-well excitons into separate periodic potentials, one can achieve multivalley (instead of $\pi$- or $0$-) condensation~\cite{Sun:2017ab}. Moreover, polariton condensation in the presence of distributed gain and loss of the single-particle states is expected to be accompanied by formation of spontaneous currents~\cite{Nalitov:2017aa}. Another significant recent finding is the flat band condensation in 1D~\cite{Baboux:2016aa} and 2D periodic systems~\cite{Klembt:2017aa,Whittaker2018cd}. Such condensates exhibit strong enhancement of the effects of polariton-polariton interaction due to the reduced kinetic energy of the particles. 

In this Letter, we consider a 1D polariton system in a complex-valued (later \textit{complex}) periodic potential and account for polariton-polariton interaction and gain saturation nonlinearity. 
We show that for the detailed description of the system, it is necessary to consider the imaginary part of the periodic potential, which describes the distributed gain and losses~\cite{Winkler:2016aa} of single-particle states in the microcavity. 
By carefully choosing the parameters of the complex potential (such as height and width of its imaginary part), we can control the state of the system and demonstrate that several conceptually different situations are possible.

In the case of relatively large width of the miniband (large energy difference between the $0$-  and $\pi$-states of the single-particle spectrum), we find that the condensate transforms from $0$-state to $\pi$-state or vice versa with the increasing interaction between the particles. This result is counterintuitive, as particles accumulate in a state which does not correspond to minimal losses. Moreover, in the presence of strong polariton-polariton repulsion, the total occupation number of the condensate is not maximized.

The crossover from 0- to $\pi$-condensate (or vice versa) happens by formation of propagating dark solitons, which is another surprising result. 
Instead of usual condensation that tends to maximize the number of particles in a single quantum state, the particles might quasi-homogeneously distribute along the dispersion curve, at certain magnitude of effective interaction between the particles, comparable with the band width. 
In this case, polaritons occupy the band more or less uniformly, and  short correlations in space and time manifest intermittency of such states.

%
%
%
\newpage

\emph{Theoretical model.---} We study the solutions of 1D Gross-Pitaevskii equation,
\begin{equation}\label{GPeq}
    \mi\hbar\partial_t\psi = 
    -\frac{\hbar^2}{2 m^*}\partial^2_x\psi + V\left(x\right)\psi 
    +\left(\alpha-\mi\beta\right)|\psi|^2\psi,
\end{equation}
where $\psi\left(x,t\right)$ is the wave function of polariton condensate, $V(x)=V(x+a)$ is the complex periodic potential with the lattice period $a$, $m^*$ is the polariton effective mass, $\alpha$ is the polariton-polariton interaction constant, and $\beta>0$ accounts for the gain-saturation nonlinearity of the system~\cite{Keeling:2008js, karpov2015dissipative}. 
Note that by scaling the wave function $\psi\rightarrow\psi/\sqrt{\beta}$, one can set $\beta=1$, obtaining the dimensionless interaction constant $\alpha/\beta$.  

\begin{figure}[t]
\centering
\vspace{-0.5cm}
\subfloat[$\Lambda\Lambda$-type]{
\hspace{-0.05\columnwidth}
\includegraphics[clip,width=0.53\columnwidth]{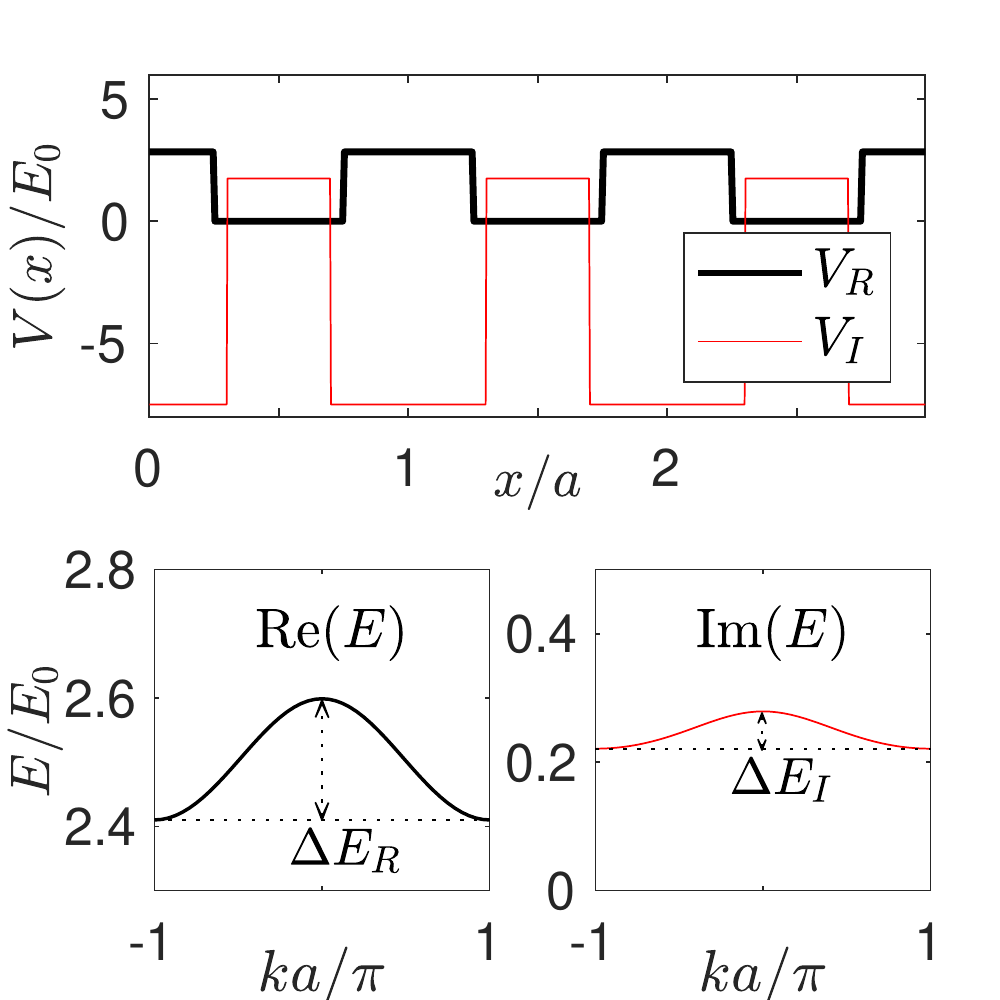}
}
\subfloat[VV-type]{
\hspace{-0.05\columnwidth}
\includegraphics[clip,width=0.53\columnwidth]{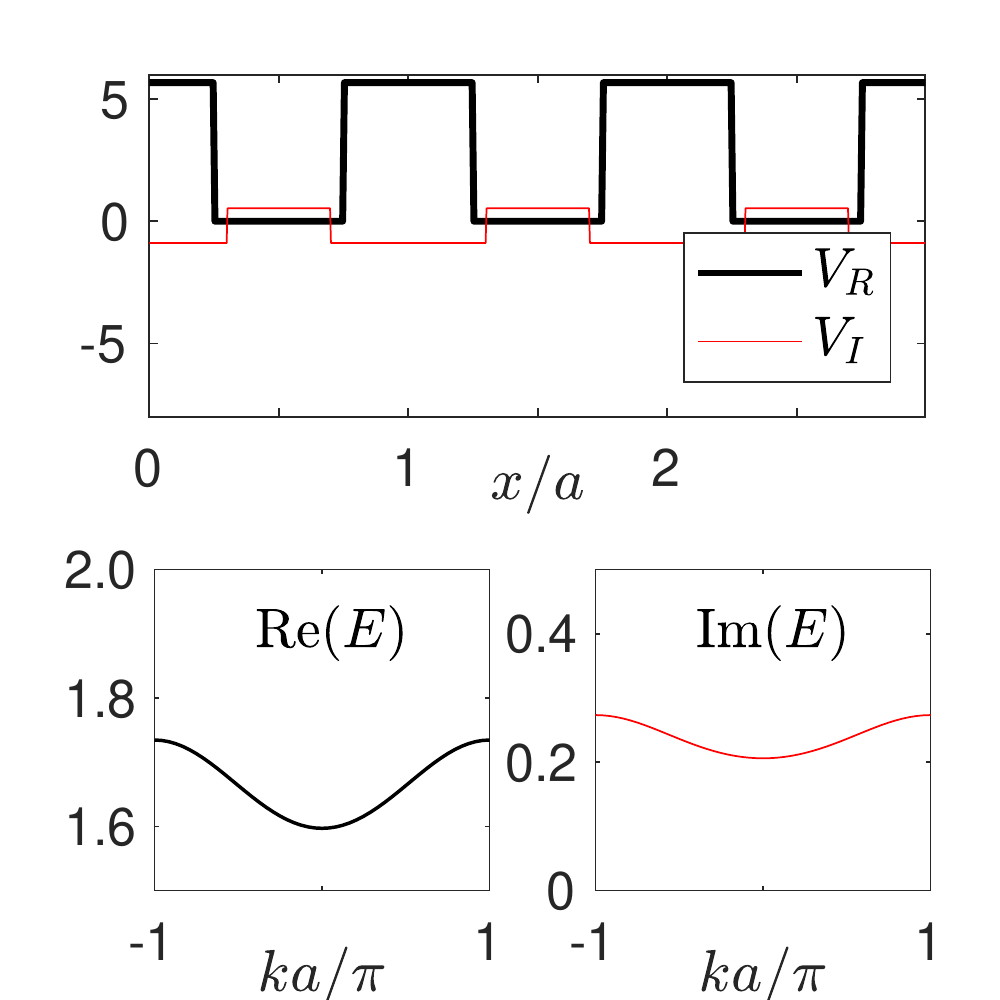}
}
\vspace{-0.1cm}
\caption{Complex potentials and dispersions of the first miniband. The real parts are plotted with thick black lines; the imaginary parts are plotted with thin red lines. (a) $\Lambda\Lambda$ case. $U = 14(\hbar^2/m^* a^2)$, $W = 3.26~U$, $\Gamma=2.64~U$. (b) VV case. $U=28(\hbar^2/m^* a^2)$, $W=0.25~U$, $\Gamma= 0.16~U$. The energy bandwidth ratio ${\Delta}E_R/{\Delta}E_I \approx 3$ (a) and $2$ (b). In both cases, $a_R = 0.5a$  and $a_I = 0.4a$. The energies are measured in the units of $E_0=\pi^2\hbar^2/2m^*a^2$.}
\label{fig:1-1}
\end{figure}

We describe the complex potential $V(x)=V_R(x)+{\mi}V_I(x)$ as a superposition of square wells in both real and imaginary part of it, but with different widths. Namely, within the unit cell, $0 \le x < a$, we have (see Fig.~\ref{fig:1-1}, upper panels)
\begin{subequations}\label{CoxV}
\begin{align}
  V_R(x) &= U\,\Theta\!\left(\left | x -\frac{a}{2} \right |-\frac{a_R}{2} \right), \\
  V_I(x) &= W\,\Theta\!\left(\frac{a_I}{2}-\left | x -\frac{a}{2} \right | \right)-\Gamma,
\end{align}
\end{subequations}
where $\Theta(x)$ is the Heaviside step function, $U$ is the height of potential barriers, $W$ describes the local gain, and $\Gamma$ defines the uniform losses in the system (due to finite polariton lifetime in the microcavity wire). Parameters $a_R$ and $a_I$ are the widths of real potential wells and imaginary potential barriers, respectively. 
Complex potential~\eqref{CoxV} reflects the experimental situation when the system is pumped from the excitonic reservoirs created in the barriers. Due to exciton-polariton repulsion, the particles move into the wells (similar to the case of 2D lattices of trapped polariton condensates~\cite{Ohadi:2017aa}), so that the gain part of the potential (parameter $W$) is located at the wells. The excitonic reservoirs also increase the barrier heights. 
It should be noted that Eq.~\eqref{GPeq} can describe the experimental set-up in which there is no real part of the potential along the wire ($V_R(x)=0$), but the incoherent pumping is periodic. In this case the pumping both creates the potential barriers and selectively pumps the wells. This set-up allows to fine tune the parameters of the potential and its period $a$. 

It is important that depending on the values of the parameters of potential~\eqref{CoxV}, the single-polariton spectrum (obtained setting $\alpha=\beta=0$ in Eq.~\eqref{GPeq}) can be of four qualitatively different types. We classify them as $\Lambda\Lambda$ [see Fig.~\ref{fig:1-1}(a), lower panel], VV [Fig.~\ref{fig:1-1}(b), lower panel], V$\Lambda$ and $\Lambda$V, depending on the position of the minimum of the energy (the real part of the eigenvalue) and the position of the minimum of the gain (the imaginary part of the eigenvalue) for the first miniband. 
For example, $\Lambda\Lambda$ type corresponds to the case when the minimum energy and minimum gain are both at the edge of the first Brillouin zone at $k=\pm \pi/a$ [see Fig.~\ref{fig:1-1}(a)]. We also can define effective widths of the bands, $\Delta E_R$ and $\Delta E_I$.

In what follows, we will consider the formation of the polariton condensate near the threshold, when the losses in the system, governed by the parameter $\Gamma$, are big enough, and the first miniband only possesses positive imaginary part of the eigenvalue, so that the particles are expected to condense into this miniband. The complex potential in Fig.~\ref{fig:1-1} is then chosen by the following two principles: 
(i) Detuning the width and depth of the well to get the ground state dispersion with the $\Lambda\Lambda$ (or VV type), and (ii) changing the magnitude of the overall shift for the imaginary part, $\Gamma$, we make the first miniband to be the only band with positive imaginary part of the eigenvalues.


\emph{Spatiotemporal dynamics of the condensate.---} For noninteracting polaritons one expects to obtain their condensate in the state with the maximum gain. Thus the number of particles is maximized in this state and it is stabilized by finite gain-dissipation parameter $\beta$. Therefore, we expect the $0$-state condensation in $\Lambda\Lambda$ and V$\Lambda$ cases, and $\pi$-state condensation in $\Lambda$V and VV cases. Numerical solutions of Eq.\ \eqref{GPeq} show that this scenario remains valid even in the presence of strong repulsion between polaritons in the V$\Lambda$ and $\Lambda$V cases. However, the polariton-polariton interaction has dramatic effect on the condensate loaded in the $\Lambda\Lambda$ and VV minibands, leading to fundamental reconstruction of the condensate state with the increase of interactions. Therefore we will mostly concentrate on the VV and $\Lambda\Lambda$ configurations.

\begin{figure}[t]
\centering
\subfloat{
\hspace{-0.50\columnwidth}
\includegraphics[clip,width=0.55\columnwidth]{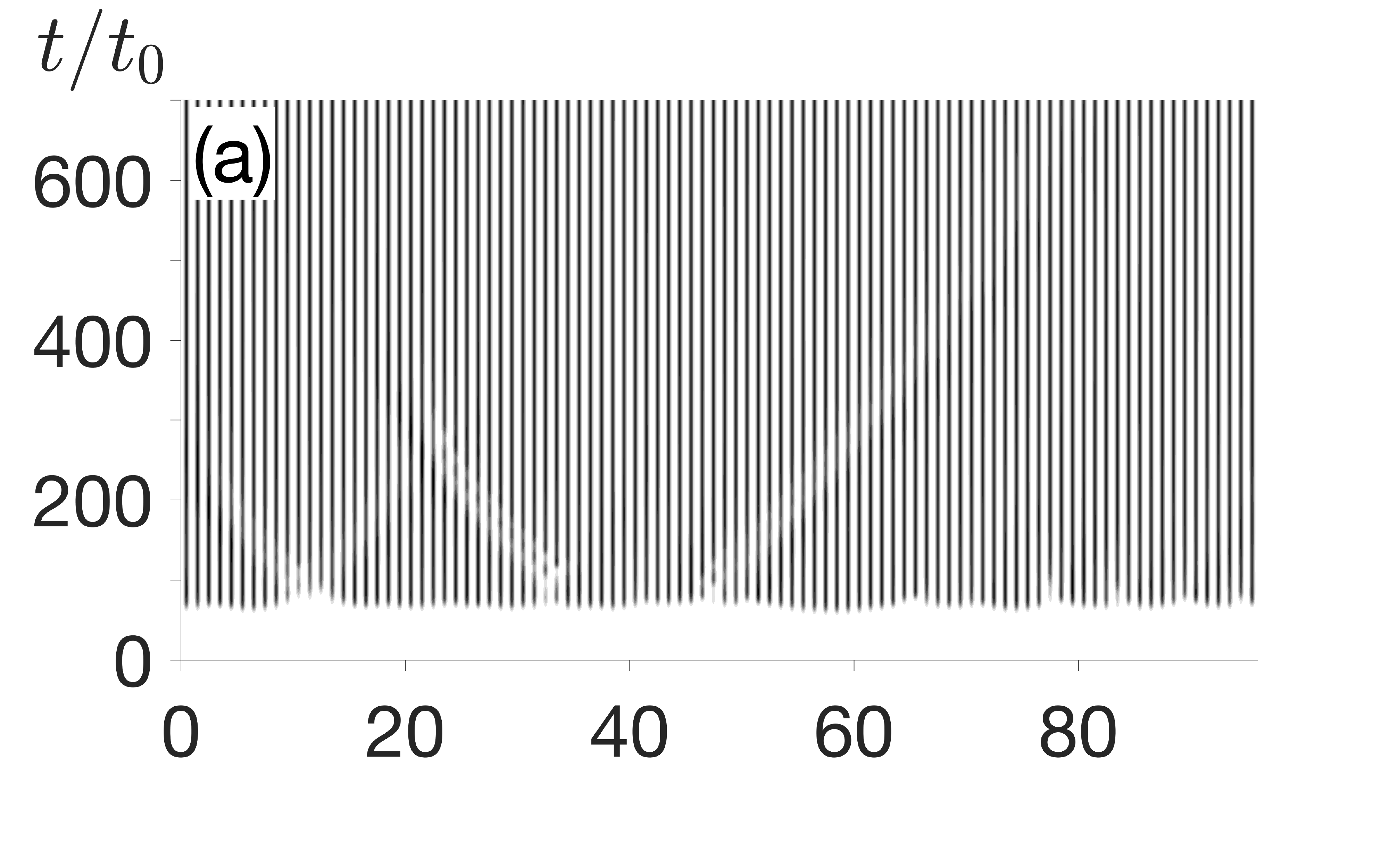}
\hspace{-0.05\columnwidth}
\includegraphics[clip,width=0.55\columnwidth]{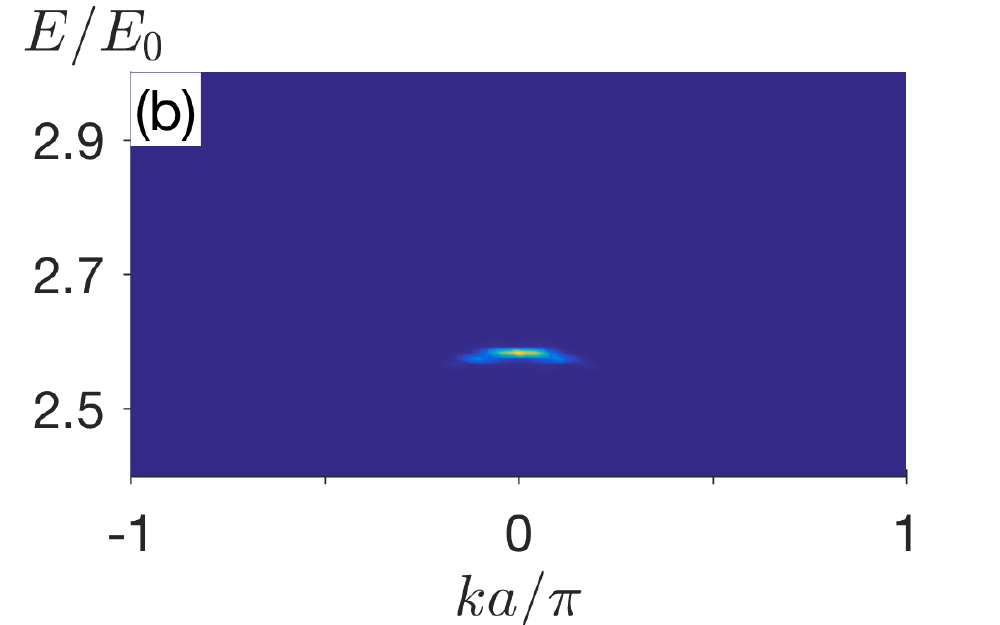}
\hspace{-0.50\columnwidth}
}\\
\vspace{-1.0cm}
\subfloat{
\hspace{-0.50\columnwidth}
\includegraphics[clip,width=0.55\columnwidth]{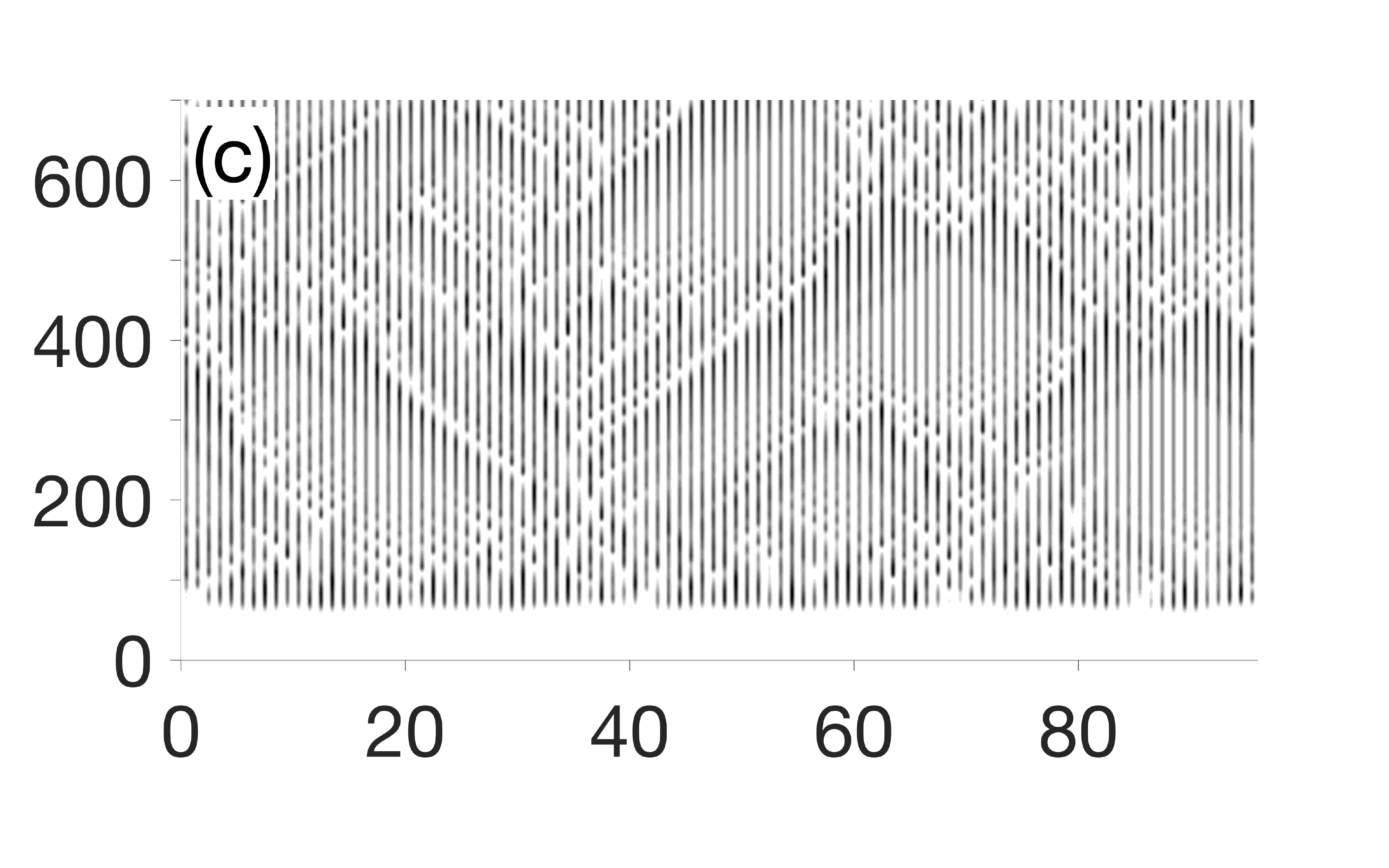}
\hspace{-0.05\columnwidth}
\includegraphics[clip,width=0.55\columnwidth]{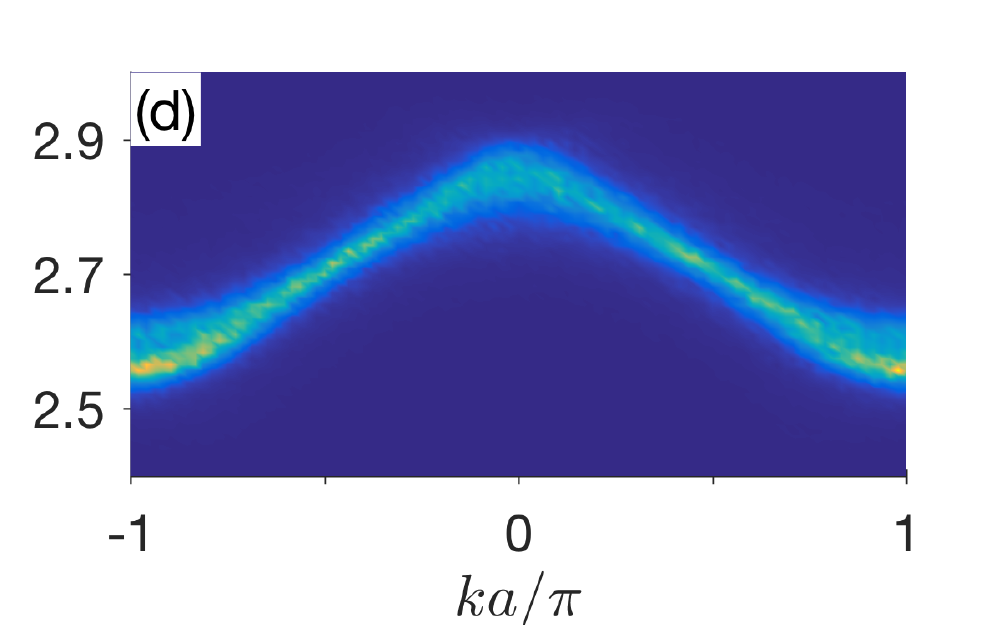}
\hspace{-0.50\columnwidth}
}\\
\vspace{-1.0cm}
\subfloat{
\hspace{-0.50\columnwidth}
\includegraphics[clip,width=0.55\columnwidth]{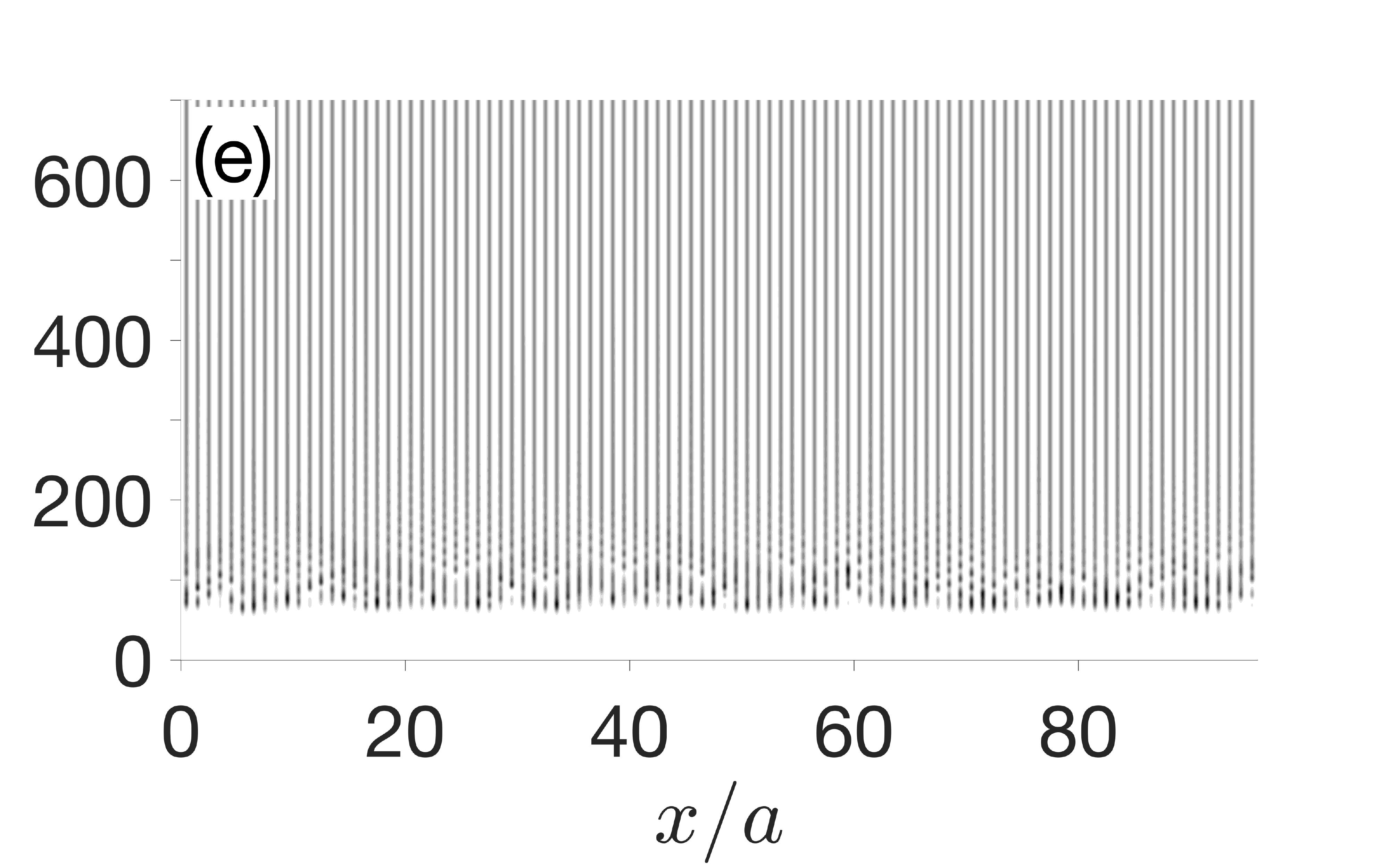}
\hspace{-0.05\columnwidth}
\includegraphics[clip,width=0.55\columnwidth]{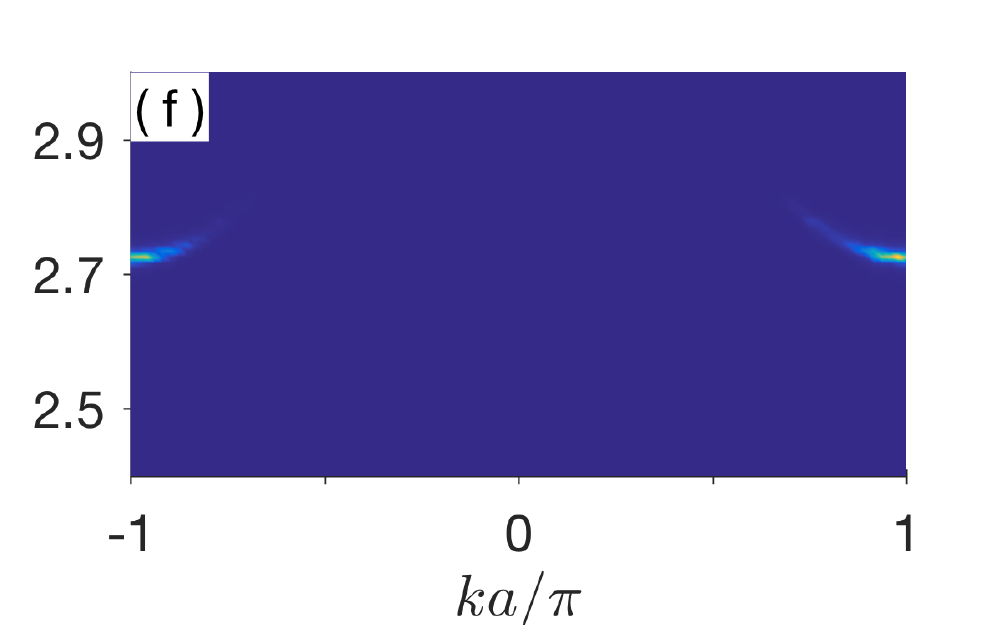}
\hspace{-0.50\columnwidth}
}
\caption{Spatiotemporal pattern of the condensate density $|\psi(x,t)|^2$ (a, c, e) and Intensity $|\psi(k,E)|^2$ (b, d, f) for $\Lambda\Lambda$ case. $\alpha/\beta = 0$ (a, b), $2$ (c, d), and $6$ (e, f). The intensity plot is obtained from averaging over 50 trajectories of different random initial noise. Time is measured in units of $t_0 \equiv \hbar/E_0$.}
\label{fig:2-1}
\end{figure} 

\begin{figure}[htbp]
\centering
\subfloat{
\hspace{-0.50\columnwidth}
\includegraphics[clip,width=0.55\columnwidth]{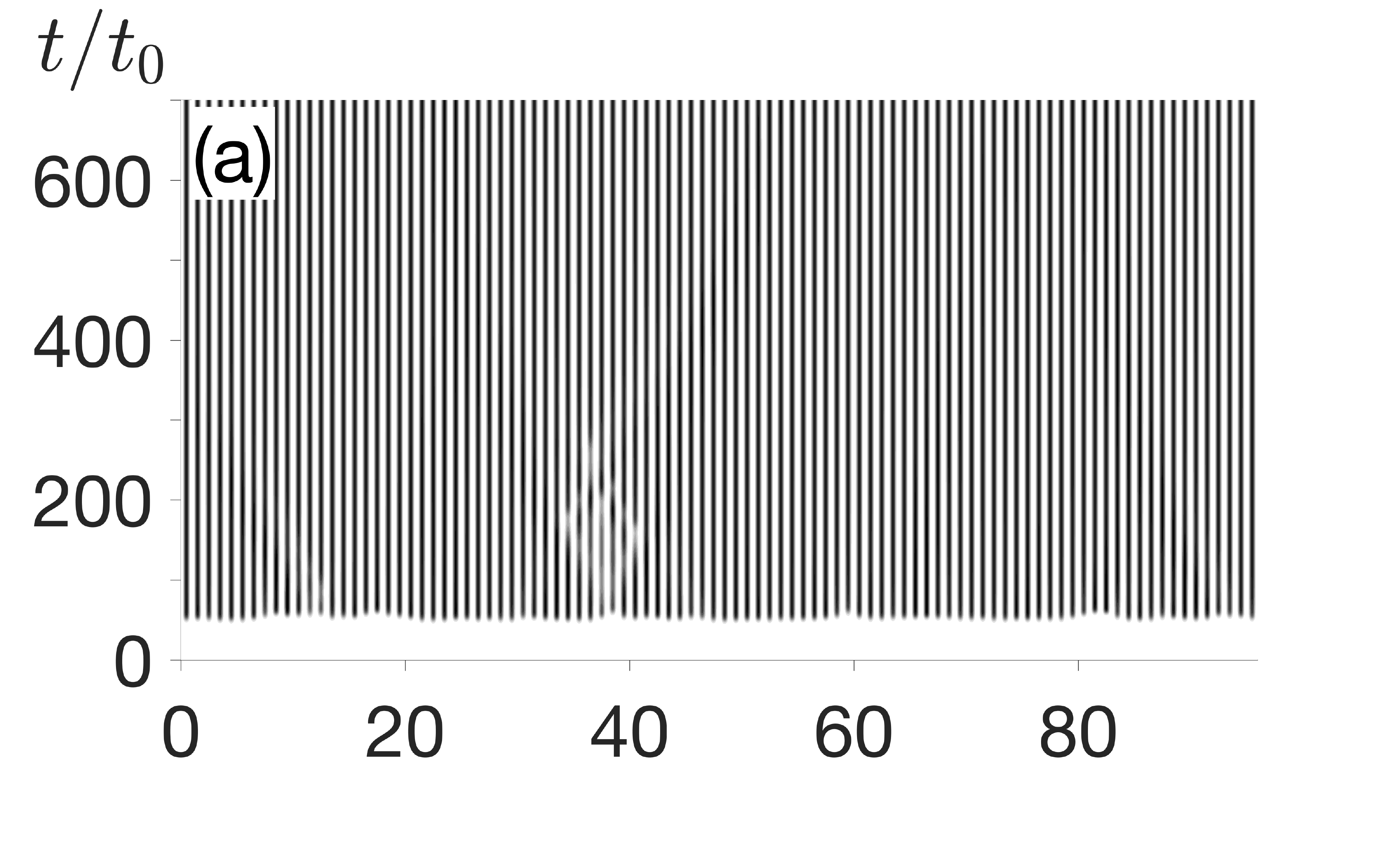}
\hspace{-0.05\columnwidth}
\includegraphics[clip,width=0.54\columnwidth]{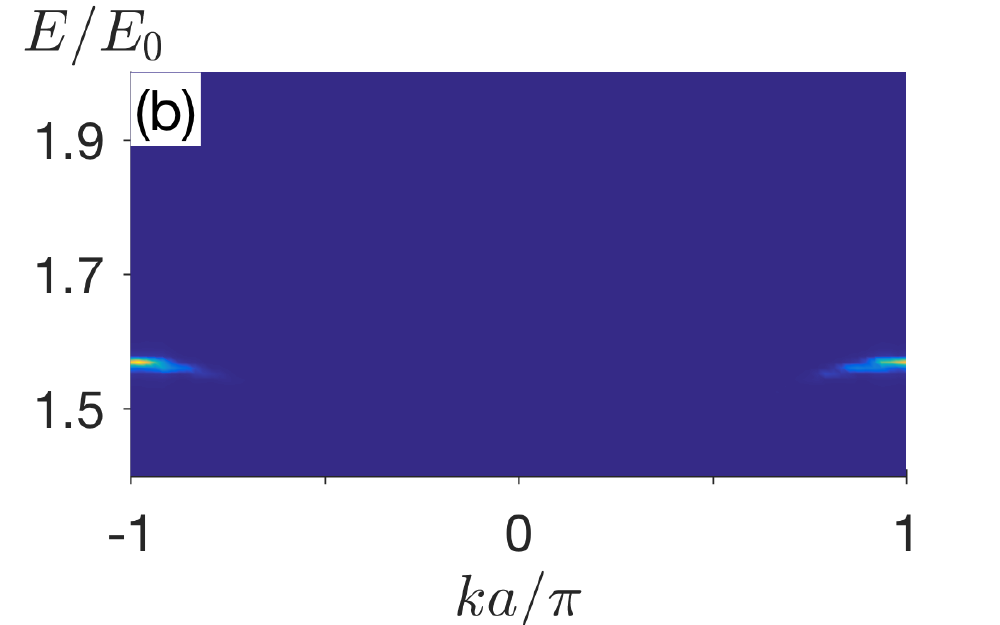}
\hspace{-0.50\columnwidth}
}\\
\vspace{-1.0cm}
\subfloat{
\hspace{-0.50\columnwidth}
\includegraphics[clip,width=0.55\columnwidth]{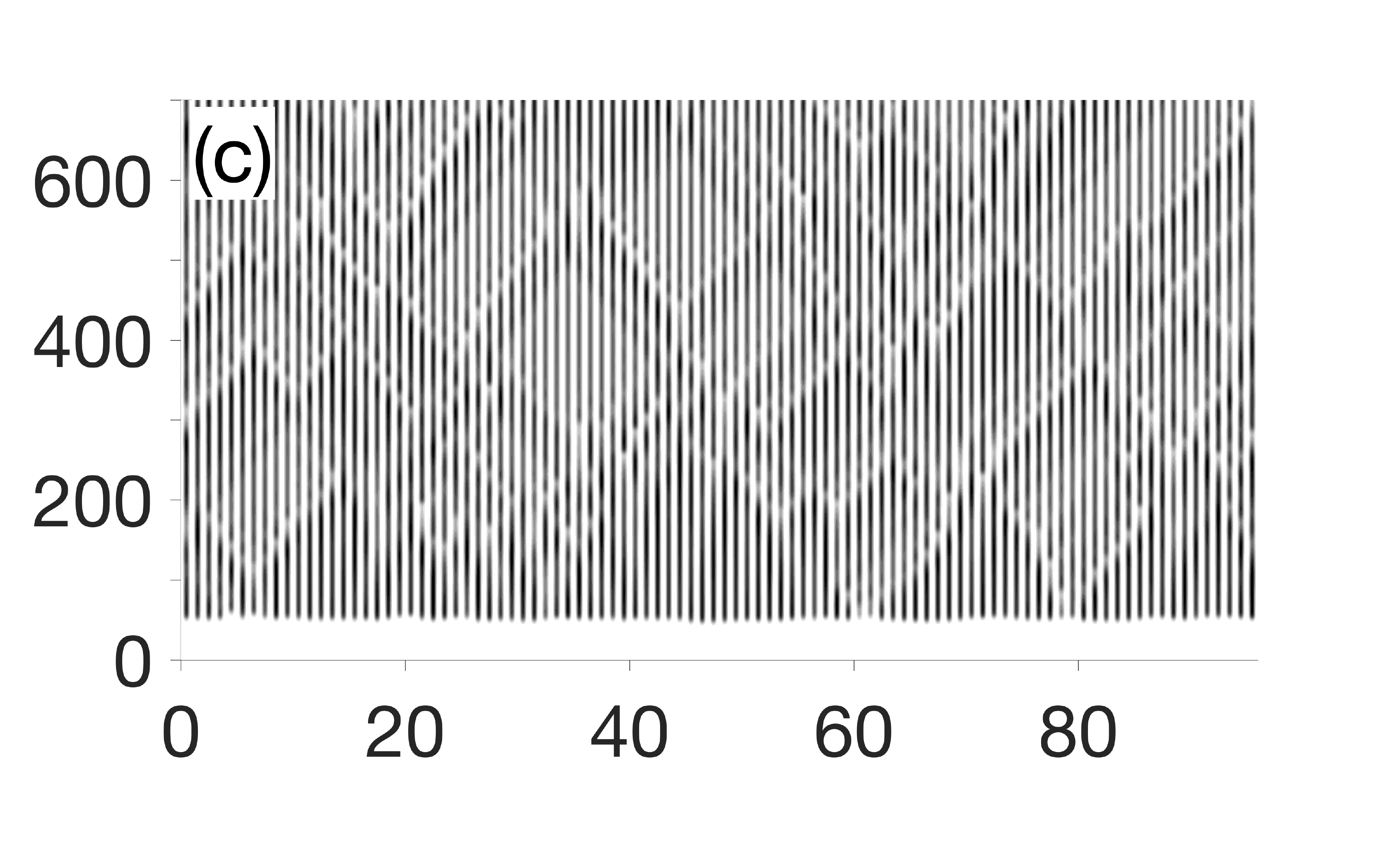}
\hspace{-0.05\columnwidth}
\includegraphics[clip,width=0.54\columnwidth]{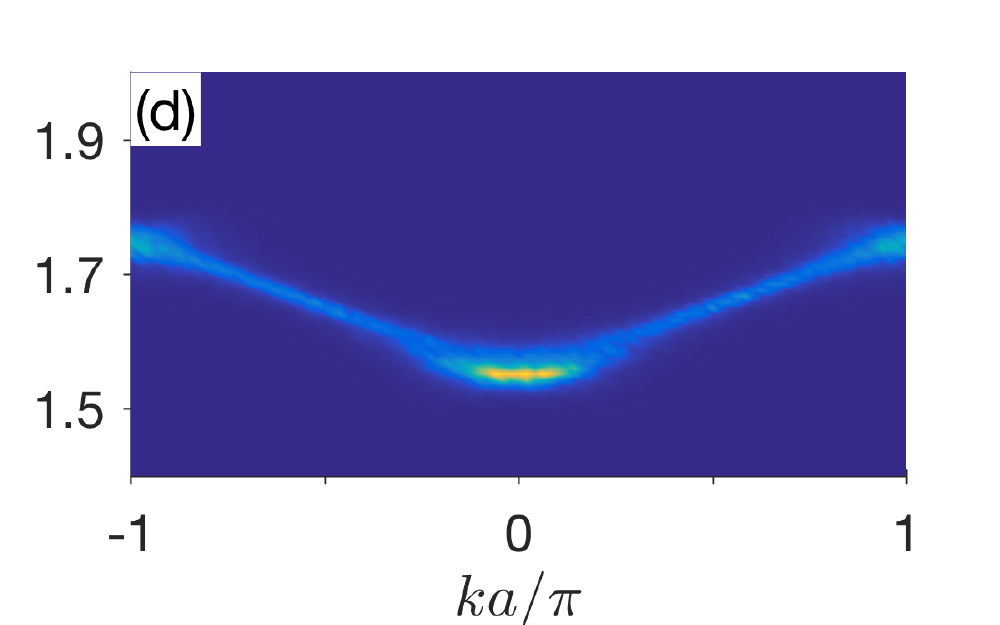}
\hspace{-0.50\columnwidth}
}\\
\vspace{-1.0cm}
\subfloat{
\hspace{-0.50\columnwidth}
\includegraphics[clip,width=0.55\columnwidth]{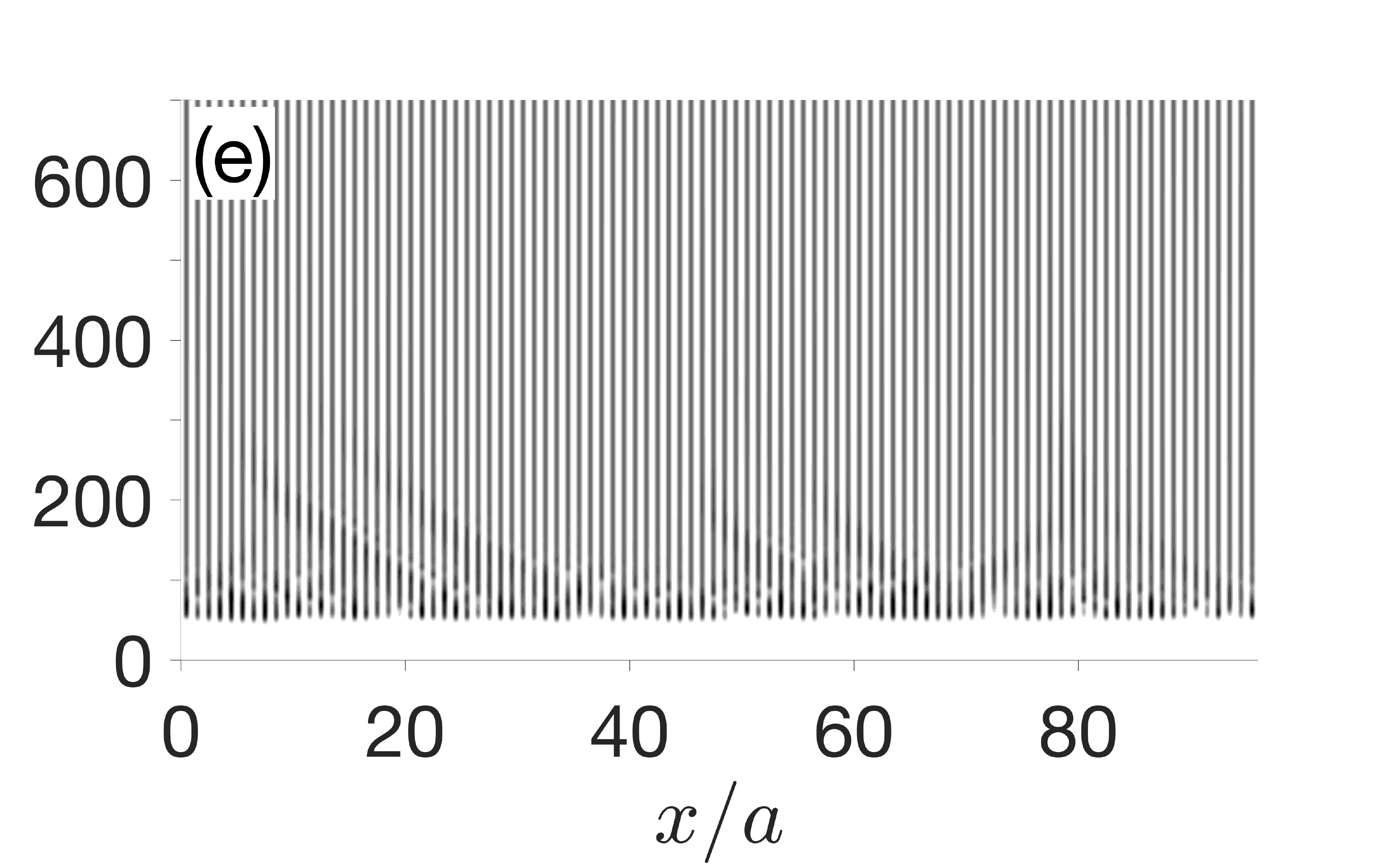}
\hspace{-0.05\columnwidth}
\includegraphics[clip,width=0.54\columnwidth]{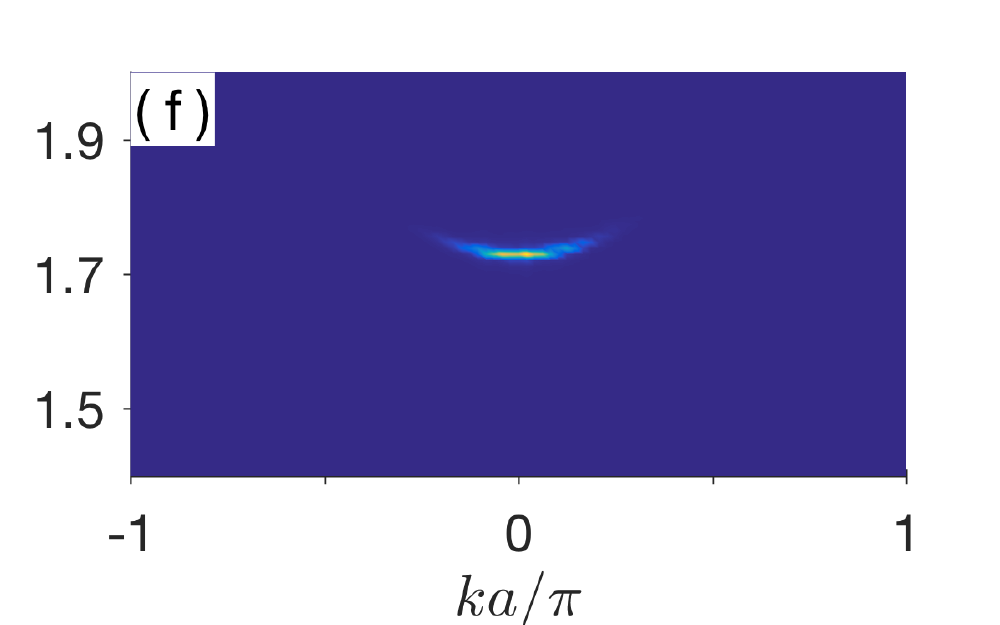}
\hspace{-0.50\columnwidth}
}
\caption{Spatiotemporal pattern of the condensate density $|\psi(x,t)|^2$ (a, c, e) and Intensity $|\psi(k,E)|^2$ (b, d, f) for VV case. $\alpha/\beta = 0$ (a, b), $0.8$ (c, d), and $2$ (e, f). Each intensity plot (b, d, f) is obtained by averaging over 50 trajectories of different random initial noise.
}
\label{fig:2-2}
\end{figure}

\begin{figure}[!t]
\centering
\subfloat{
\hspace{-0.10\columnwidth}
\includegraphics[clip,width=0.50\columnwidth]{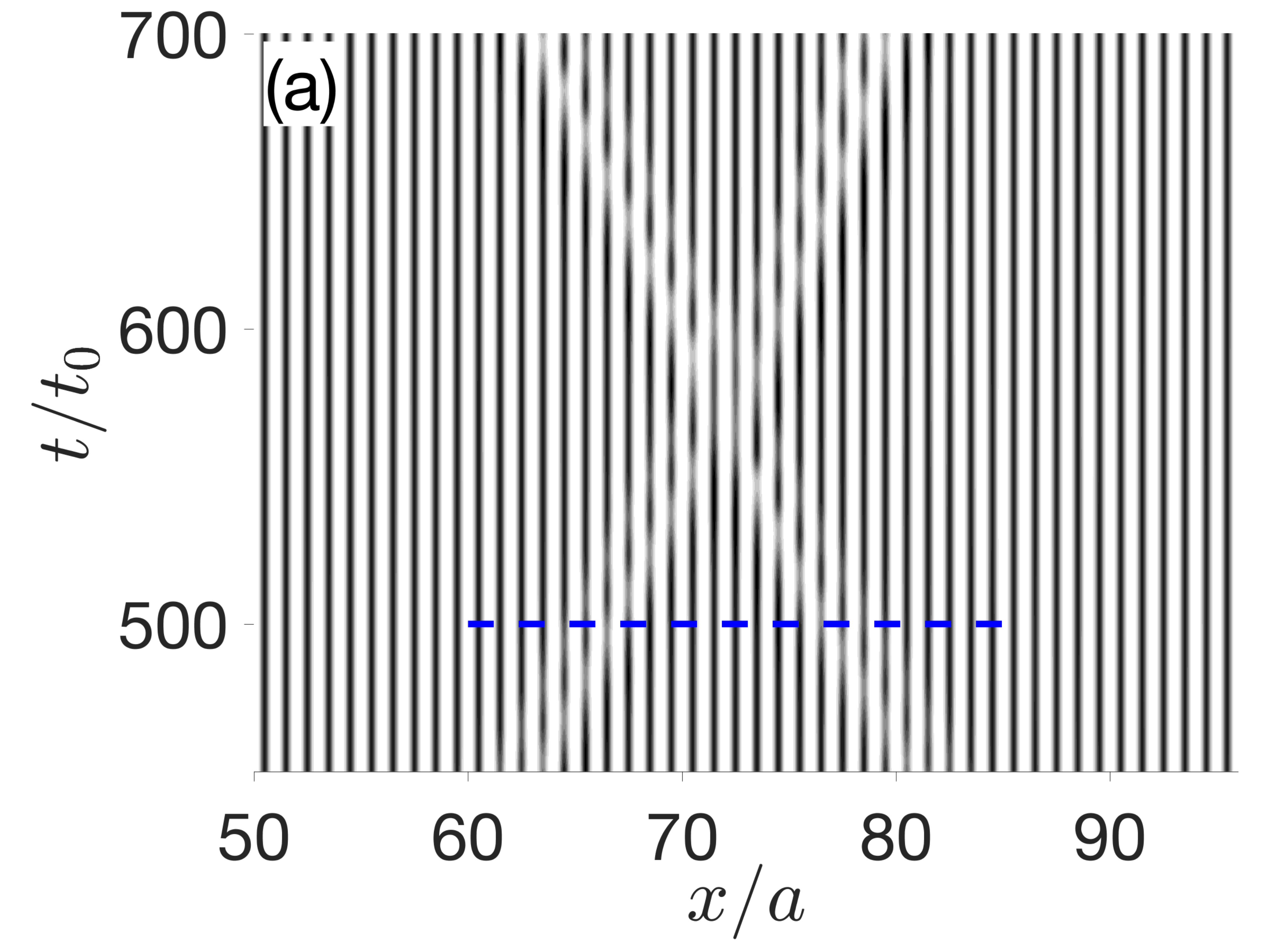}
\hspace{-0.02\columnwidth}
\includegraphics[clip,width=0.50\columnwidth]{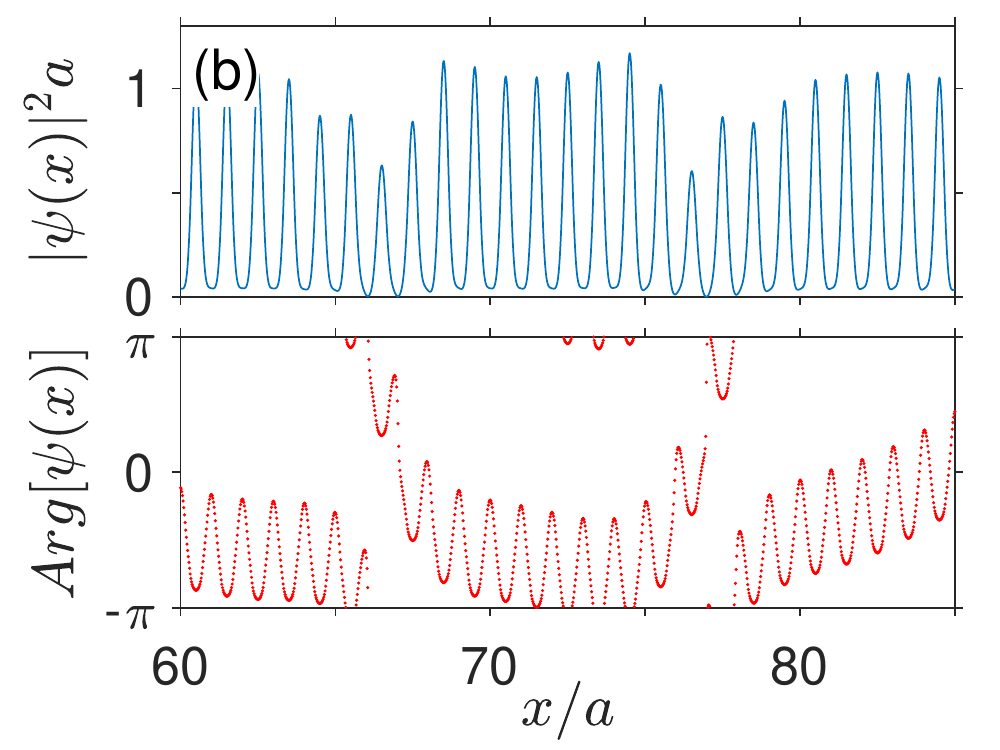}
\hspace{-0.10\columnwidth}
}\\
\vspace{-0.2cm}
\subfloat{
\hspace{-0.10\columnwidth}
\includegraphics[clip,width=0.50\columnwidth]{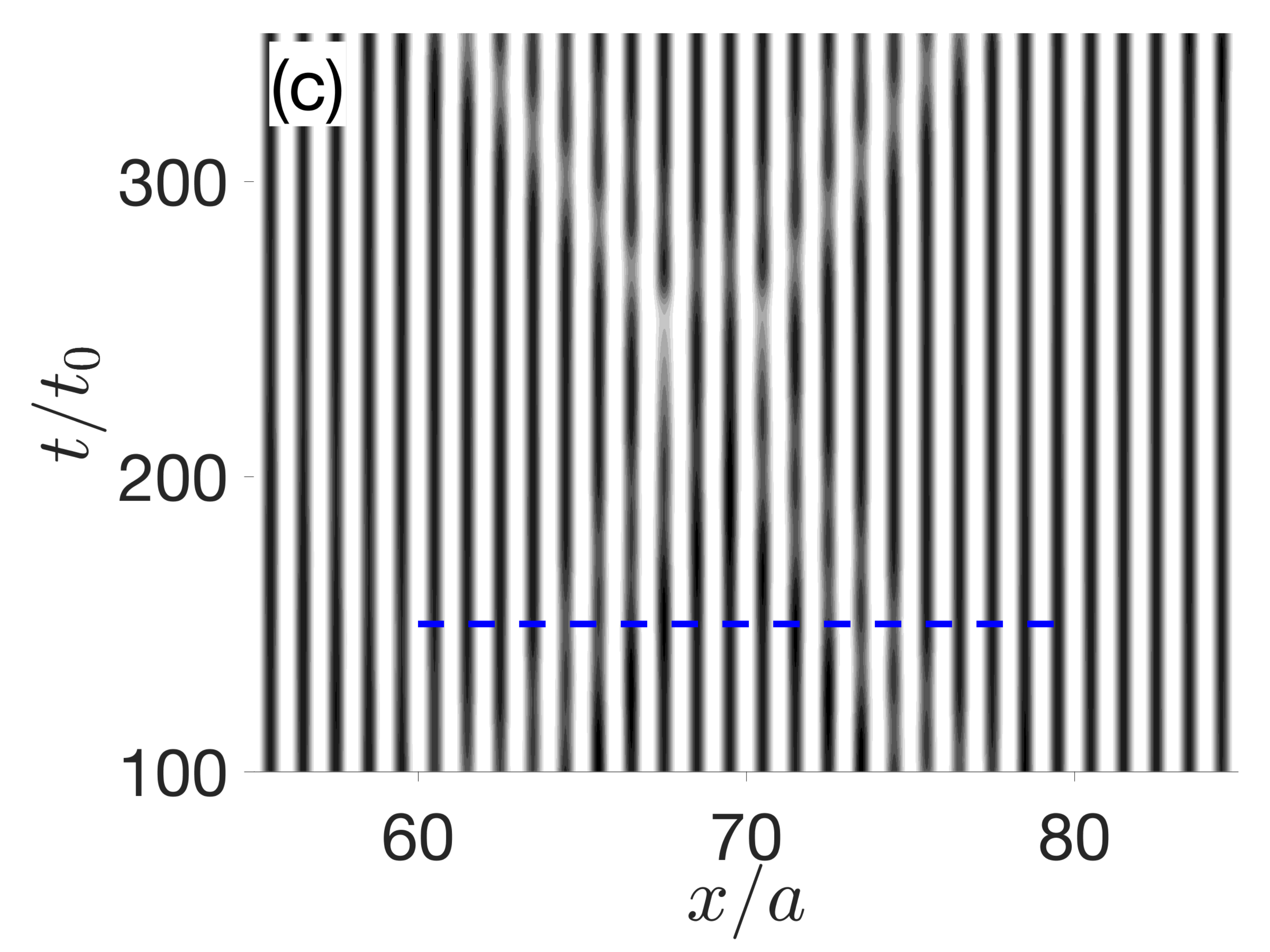}
\hspace{-0.02\columnwidth}
\includegraphics[clip,width=0.50\columnwidth]{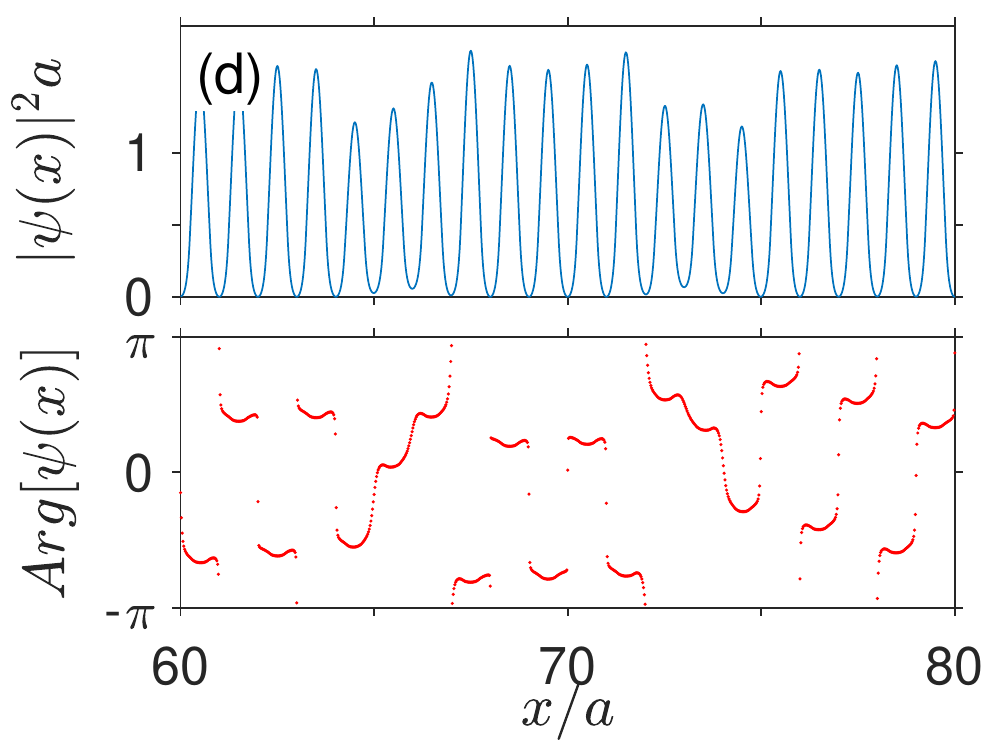}
\hspace{-0.10\columnwidth}
}

\caption{Propagation and collision of a pair of solitons in the $\Lambda\Lambda$ case (a,b) and VV case (c,d) for $\alpha/\beta=0.4$, when the number of solitons is small. (b,d) The upscaled profiles of the particle density and the phases of the wave functions just before the collision [along the thick dashed blue lines in panels (a,c), respectively].}
\label{fig:4}
\end{figure}

Figures~\ref{fig:2-1} and~\ref{fig:2-2} show spatiotemporal dynamics of the condensate for the $\Lambda\Lambda$ and VV cases. The left-hand-side (l.h.s.) columns of these figures show spatiotemporal patterns of the condensate density $|\psi(x,t)|^2$ and the right-hand-side (r.h.s.) columns show the polariton emission intensities $|\psi(k,E)|^2$ averaged over 50 trajectories with different random small initial seeds of noise. 
In density plots, shown in the panels (a), (c), and  (e), the local maxima of the condensate density (dark black vertical lines along the time axis) are at the centers of wells of the real potential $V_R$, where the maximal gain is attained (see also Fig.~\ref{fig:4}(a,b)).

Descending from panels (a,b) to panels (e,f) in Figs.~\ref{fig:2-1} and \ref{fig:2-2}, the dimensionless parameter $\alpha/\beta$ increases and we observe considerable changes in the spatiotemporal density patterns and condensation states.
(a,b) When $\alpha/\beta=0$, the condensate is formed in the state with the maximum gain: $0$-state in $\Lambda \Lambda$ case and $\pi$-state in VV case. The defects appearing at the early stage of evolution, dissipate away at later times.
(c,d) When $\alpha/\beta$ takes an intermediate value, polaritons no longer accumulate in the state with a well-defined wave vector, but rather they distribute along the whole miniband. As one can see from corresponding l.h.s. panels, strong spatiotemporal chaos is present in this case. 
(e,f) Surprisingly, with further increase of $\alpha/\beta$, the well-defined condensation takes place again, but now the condensate is formed at the minimum of the dispersion, in the state with the smallest gain. It corresponds to $\pi$-state in $\Lambda \Lambda$ case and $0$-state in VV case. This result is in contrast with the one obtained in the zero interaction regime [compare panels (b) and (f)].

The density patterns presented in panels (c) resemble those of spatiotemporal intermittency in the 1D complex Ginzburg-Landau equation (CGLE)~\cite{Chate:1994,Melo:1993,Hecke:1998},
which can be written in a form of the nonlinear Schr\"odinger equation:
\begin{equation}\label{eq:NLSE}
\mi \partial_t A = \mi A + (c_1+ \mi )\partial_x^2 A + (c_3 - \mi)|A|^2 A \ .
\end{equation}
The nonlinear term here takes the same form as in Eq.\ \eqref{GPeq} and the linear terms for $c_1>0$ correspond to the complex-valued energy dispersion of $\Lambda \Lambda$-type even though the dispersion in Fig.~\ref{fig:1-1} is not a quadratic but a periodic function. The parameters $c_1$ and $c_3$ play similar roles as ${\Delta}E_R/{\Delta}E_I$ and $\alpha/\beta$ in our system, respectively. However, the shapes of real and imaginary parts of the dispersion in Fig.~\ref{fig:1-1} are not exactly proportional to each other
and the \textit{continuous} translational symmetry of CGLE is reduced to a \textit{discrete lattice} translational symmetry due to the periodic potential.

There is also another important difference. 
Even though the offset term $\mi A$ in Eq.\ \eqref{eq:NLSE} admits only a restricted range of wave numbers ($-1<k<1$) to have a positive gain with a maximum at $k=0$, the edge points $k=\pm 1$ have {\it zero} gain, and therefore the condensate can not be formed at the edge.
Instead, the polariton system is characterized by edge points of the first Brillouin zone  $k = \pm \pi/a$ with {\it finite} gain and the singularity in the density of states. It is this feature that leads to the possibility of formation of the polariton condensate at the edge.

\begin{figure}[t]
\centering
\subfloat{
\hspace{-0.50\columnwidth}
\includegraphics[clip,width=0.55\columnwidth]{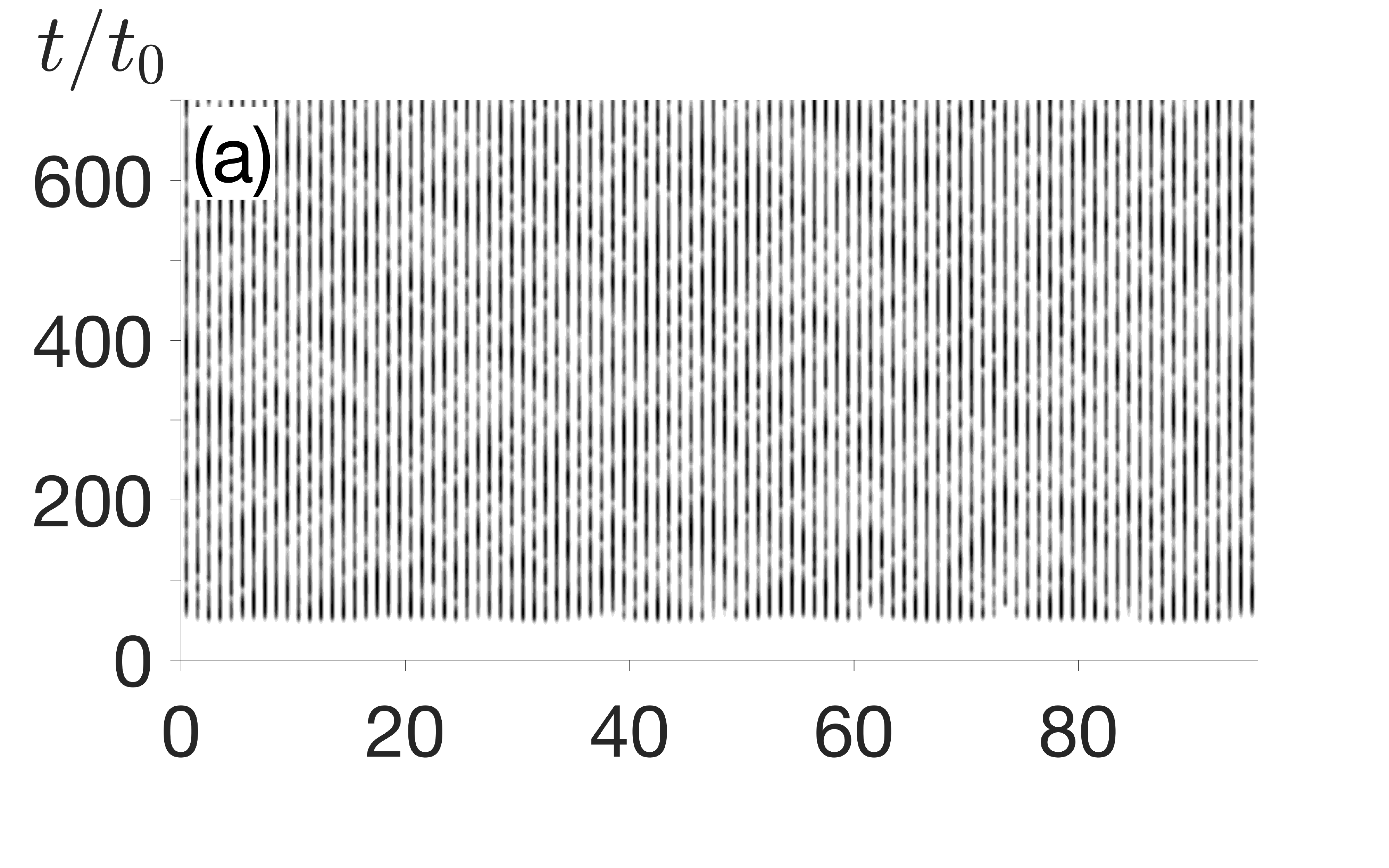}
\hspace{-0.05\columnwidth}
\includegraphics[clip,width=0.55\columnwidth]{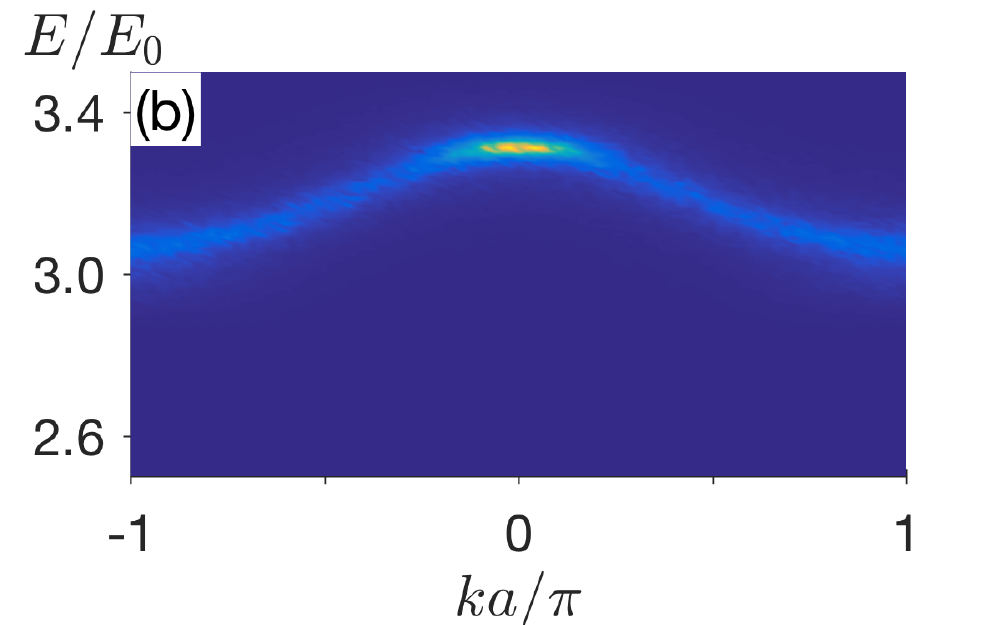}
\hspace{-0.50\columnwidth}
}\\
\vspace{-1.0cm}
\subfloat{
\hspace{-0.50\columnwidth}
\includegraphics[clip,width=0.55\columnwidth]{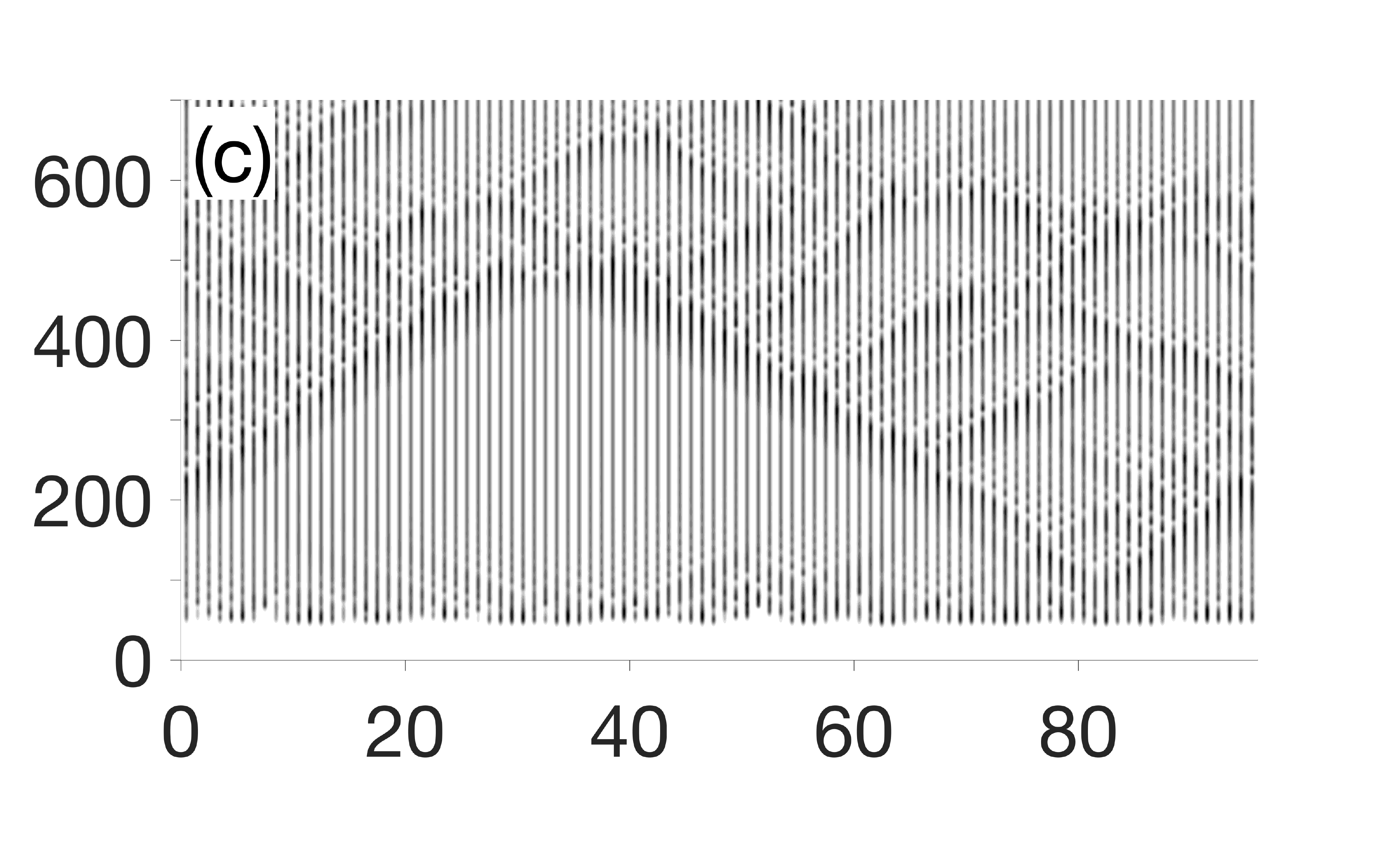}
\hspace{-0.05\columnwidth}
\includegraphics[clip,width=0.55\columnwidth]{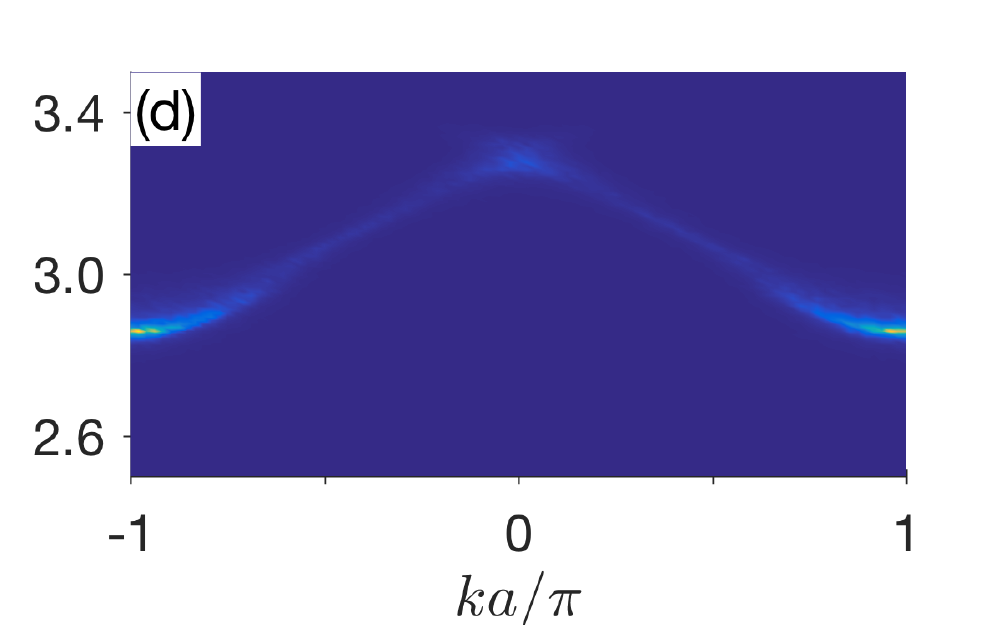}
\hspace{-0.50\columnwidth}
}\\
\vspace{-1.0cm}
\subfloat{
\hspace{-0.50\columnwidth}
\includegraphics[clip,width=0.55\columnwidth]{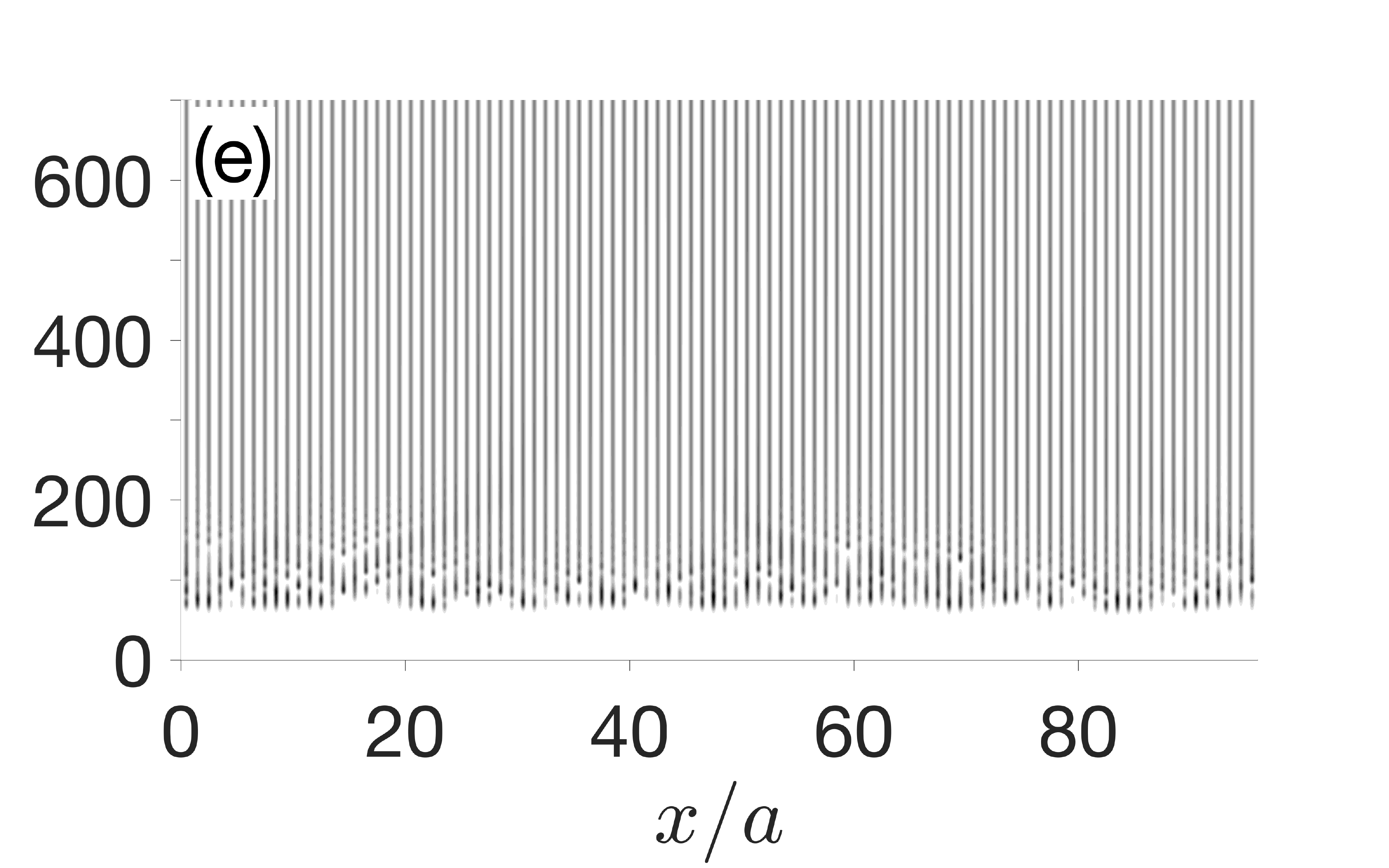}
\hspace{-0.05\columnwidth}
\includegraphics[clip,width=0.55\columnwidth]{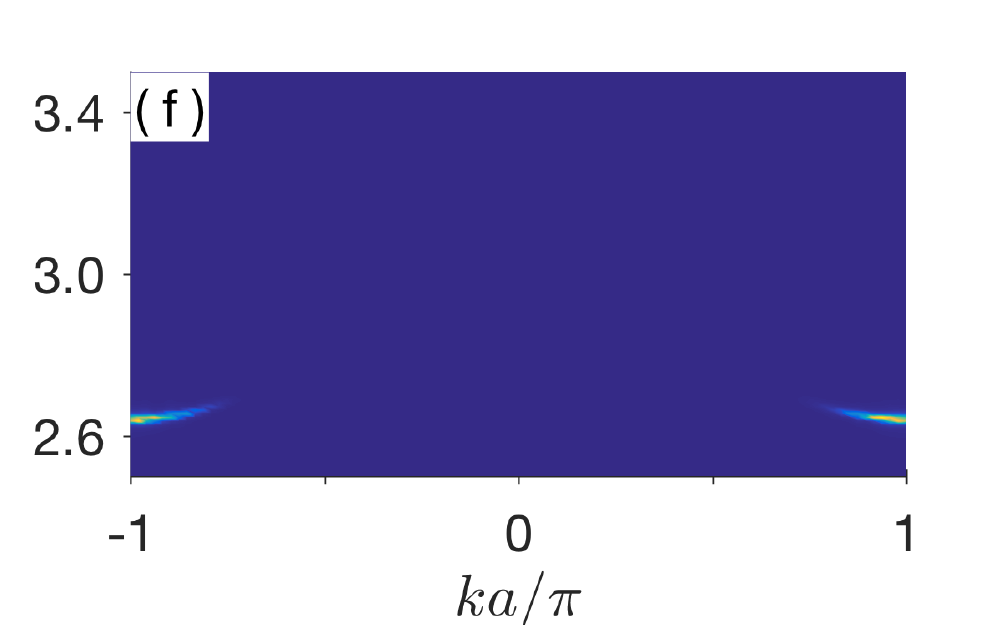}
\hspace{-0.50\columnwidth}
}
\caption{Spatiotemporal pattern of the condensate density $|\psi(x,t)|^2$ (a, c, e) and intensity $|\psi(k,E)|^2$ (b, d, f) for the $\Lambda\Lambda$ cases with $\alpha/\beta = 4$ fixed. The ratios of the real and imaginary parts of the energy bandwidths are ${\Delta}E_R/{\Delta}E_I \approx 1$, 2, and $4$ from (a, b) to (e, f). 
$W = 3.82~U$ (a,b), $3.43~U$ (c,d), and $3.18~U$ (e,f) with $U=14(\hbar^2/m^* a^2)$. 
$\Delta E_R \approx  0.107~E_0$ (a,b), $0.164~E_0$ (c,d), and $0.201~E_0$ (e,f) with $E_0=\pi^2\hbar^2/2m^*a^2$.}
\label{fig:3-1}
\end{figure}

It is known~\cite{Chate:1994} that in some parameter range of CGLE, there is a transition from a plain-wave phase to a spatiotemporal intermittent phase. In our system, we predict three different phases: $0$-, $\pi$-state Bloch-waves (instead of single plane wave) and a spatiotemporal intermittency state (which separates the $0$- and $\pi$-states).
The crossover from the spatiotemporal intermittent dynamics to the $\pi$-condensate in the $\Lambda\Lambda$ case (or to the 0-condensate in the VV case) is not always possible. The formation of new condensate phase depends not only on the polariton interaction strength $\alpha/\beta$, as has been discussed above, but also on the ratio ${\Delta}E_R/{\Delta}E_I$. 
Figure~\ref{fig:3-1} demonstrates (for the $\Lambda\Lambda$ case) that a sufficiently large value of the ratio ${\Delta}E_R/{\Delta}E_I$ is required to reach a $\pi$-state from the mixture state of the spatiotemporal intermittency.

%
%


\emph{Soliton dynamics.---} At weak polariton-polariton interaction, the long-range order of the $0$-condensate in the $\Lambda\Lambda$ case and the $\pi$-condensate in the VV case is destroyed by formation of propagating defects. Each defect extends only over a few lattice constants, and it is characterized by the suppression of the condensate occupation and by the phase slips. 

As an example, Fig.~\ref{fig:4} shows the collision events of a pair of such defects propagating towards each other. The individual collisions are seen only for weakly interacting polaritons, when the concentration of defects is small. 
Each defect is characterized by the depletion of the particle density together with an abrupt change in the phase of the wave function at two edge points of the defect, where the phase change is close to $\pi$ (instead of 0) in the $\Lambda\Lambda$ case, while it becomes less than $\pi$ in the VV case. The fact that the defects maintain their properties after the collision indicates that they can also be considered as dark solitons. We note, however, that these solitons are different from the Bekki-Nozaki hole solutions of CGLE~\cite{Bekki:1985} or dissipative Gross-Pitaevskii equation for polariton mean field~\cite{Xue:2014}. In our case, solitons represent the nuclei of a new condensate phase.

The density of solitons increases with $\alpha/\beta$, and some of them attach to each other, forming wide soliton domains, which one can see in Figs.~\ref{fig:2-1}(c) and~\ref{fig:2-2}(c). 
In $\Lambda\Lambda$ case, with further increase of polariton-polariton interaction and for sufficiently large ratio ${\Delta}E_R/{\Delta}E_I$ as in Fig.~\ref{fig:2-1}(e), a complete array of dark solitons is formed. The wave function phase changes by $\pi$ per every lattice constant and the quasi-long-range order appears again, manifesting the formation of $\pi$-condensate phase.


\emph{Conclusions.---} Interacting exciton polaritons loaded into a one-dimensional microcavity wire with a periodic potential and periodic distribution of losses can condense into nontrivial states, where losses are not minimized but maximized. Under certain conditions, polaritons can form space-time intermittency phase, which separates two condensate phases with minimal and maximal losses.   
The reconstruction of the condensate wave function takes place by proliferation of dark solitons along the periodic structure. The nuclei of the new condensate phase, which are characterized by maximization of losses, are formed with increasing polariton-polariton interaction, and they can be seen as a result of gluing the dark solitons together.

We thank Boris Altshuler, Alexey Andreanov, and Sergej Flach for useful discussions. The authors acknowledge the support of the Institute for Basic Science in Korea (Project No.~IBS-R024-D1). YGR acknowledges support from CONACYT (Mexico) under the Grant No.\ 251808.

\bibliographystyle{apsrev4-1}
\bibliography{rec}

\end{document}